# Spin Excitations in $BaFe_{1.84}Co_{0.16}As_2$ Superconductor Observed by Inelastic Neutron Scattering

D. Parshall,[1] K. A. Lokshin,[2] Jennifer Niedziela,[1,3] A. D. Christianson,[3] M. D. Lumsden,[3] H. A. Mook,[3] S. E. Nagler,[3] M. A. McGuire,[3] M. B. Stone,[3] D. L. Abernathy,[3] A. S. Sefat,[3] B. C. Sales,[3] D. G. Mandrus,[3] and T. Egami[1,2,3]

[1]*Department of Physics and Astronomy, University of Tennessee, Knoxville, TN 37996*

[2]*Department of Materials Science and Engineering, University of Tennessee, Knoxville, TN 37996*

[3]*Oak Ridge National Laboratory, Oak Ridge, TN 37831, USA*

Superconductivity appears to compete against the spin-density-wave in Fe pnictides. However, optimally cobalt doped samples show a quasi-two-dimensional spin excitation centered at the (0.5, 0.5, $L$) wavevector, "the spin resonance peak", that is strongly tied to the onset of superconductivity. By inelastic neutron scattering on single crystals we show the similarities and differences of the spin excitations in BaFe$_{1.84}$Co$_{0.16}$As$_2$, with respect to the spin excitations in the high-temperature superconducting cuprates. As in the cuprates the resonance occurs as an enhancement to a part of the spin excitation spectrum which extends to higher energy transfer and higher temperature. However, unlike in the cuprates, the resonance peak in this compound is asymmetric in energy.

The intense interest in the recently discovered Fe-based superconductors[1], which show critical temperatures up to 55K[2], stems partly from the possibility that understanding these compounds may shed light on the mechanism of high-temperature superconductivity in the cuprates. Just as in the cuprates, the superconducting phase in the Fe-based compounds is adjacent to an antiferromagnetic phase, suggesting that spin degree of freedom plays a role in the development of superconductivity[3, 4, 5, 6]. Indeed, strong magnetic excitations linked to superconductivity were observed by neutron scattering[7, 8, 9], even though long-range magnetic order is absent. The purpose of this work is to characterize the spin excitations in superconducting $BaFe_{1.84}Co_{0.16}As_2$ to higher ranges of energy transfer and temperature than has been previously reported. These data permit a detailed comparison to the spin excitations found in the cuprates.

Three single crystals of optimally doped $BaFe_{1.84}Co_{0.16}As_2$ with a total mass of ~ 1.8 g were co-aligned in the HHL plane[8]. Inelastic neutron scattering measurements were performed on the HB-3 triple axis spectrometer at the HFIR and the ARCS time-of-flight chopper spectrometer at the SNS, both at Oak Ridge National Laboratory. The HB-3 measurements were performed using pyrolitic graphite crystals for the monochromator and analyzer, operating at a fixed final energy of 30.5 meV. The collimation was set at 48´-40´-80´-120´, producing an energy resolution of 3 meV at the elastic position. To reduce higher order wavelength contamination, a pyrolitic graphite filter was placed after the sample. To reduce spurious signals caused by aluminum granularity, the sample can used for the measurements at the HFIR was constructed using vanadium sheet metal. For the ARCS measurements an incident energy of 60 meV was used, with an energy resolution of about 2 meV at the elastic position. The sample was configured to have the (0, 0, 1) and (1, 1, 0) axes in the scattering plane. This configuration confined the momentum exchange vector, $Q$, to the plane defined by $(H, H, L)$ in reciprocal lattice units for the HB-3 measurements. For the ARCS measurements, some range in $K$ for $Q = (H, K, L)$ was measured by the two-dimensional detector banks. The ARCS data analysis was performed using the *mslice* program.

First we focus on the part of the scattering intensity measured by HB-3 that appears only below $T_C$ (= 22 K), which constitutes the resonance peak[7, 8, 9]. Figure 1 shows the difference between the data taken at two temperatures, $I(T = 16 K) - I(T = 30 K)$, at $Q = (0.5, 0.5, 0)$ and $(0.5, 0.5, 2)$. In the previous work, the energy range was kinematically restricted to 12 meV because of the choice of the final energy at 14.5 meV. Here, the range has been extended to higher excitation energies by the use of a higher final energy, and also by looking at wavevectors with $L = 2$. However, spurious peaks involving elastic scattering from the sample and inelastic scattering from the analyzer or the monochromator limited the effective energy transfer range to 6 to 17 meV for $L = 0$ and 7 to 24 meV for $L = 2$. The data taken for $L = 0$ and 2 overlap well. The shape of the excitation spectrum thus determined is not a simple Gaussian, but is rather asymmetric. It rises sharply from $E = 6$ meV to a peak at $E = 8.5$ meV, then slowly decreases to zero at approximately 18-19 meV. In the cuprates this peak is usually narrower and more symmetric in energy, when energy is normalized by the peak energy[10, 11].

Even though these measurements were made at low values of $Q$, the phonon dispersion could interfere with the experiment. In order to rule out this possibility, a measurement was made at $Q$ = (2.5, 2.5, 0). Again the difference in the data taken at two temperatures, $I(T = 16$ K$) - I(T = 30$ K$)$, is shown in Fig. 2. This is an equivalent symmetry point to $Q$ = (0.5, 0.5, 1), but since magnetic scattering falls off rapidly with $Q$, it should have a much reduced magnetic contribution. At the same time since the phonon intensity is proportional to $Q^2$, the intensity of the phonon with in-plane polarization should be enhanced by the factor of 25 compared to the data in Fig. 1. It is obvious that this contribution is negligibly small, and the observed intensity shown in Fig. 1 is almost totally due to spin excitations.

The difference in the intensities for the $L$-scans at $H = K = 0.5$ and $E = 12$ meV, measured at $T = 16$ K and at 30 K, is shown in Figure 3. The intensity shown here represents approximately 10% of the intensities at each temperature before subtraction. Two high points in the intensity around $L = 1.2$ are most likely due to statistical fluctuations. The excitation spectrum shown here has almost no dependence on $L$, indicating the two-dimensional nature of the excitation. The weak L-dependence observed here at an energy transfer of 12 meV, and previously at an energy transfer of 9.5 meV[8], is consistent with the expected small changes in the magnetic form factor over this range, as shown by the solid line.

The excitation in general appears sharply peaked in $H, K$ at [0.5, 0.5], as shown earlier[8]. We determined the width of the temperature difference in the excitation spectrum, $I(T =$ 16K) - $I(T = 30K)$, at $E = 10$ meV for $L = 0$, and for $E = 12$ and 14 meV for $L = 2$. The peak was fit to a single Gaussian; the FWHM was found to be ~ 0.16 Å$^{-1}$ for $E = 10$ meV, ~ 0.20 Å$^{-1}$ for $E = 12$ meV, and ~ 0.27 Å$^{-1}$ for $E = 14$ meV. The total integrated intensity changed very little. These increases in widths are consistent with the previous observation[8].

So far we have focused on the increase in the scattering intensity when going below $T_C$. However, the scattering intensity persists above $T_C$, as shown earlier[8]. Both above and below $T_C$, the excitation extends to higher energy, as can be seen in the ARCS data. Fig. 4 shows the two-dimensional excitation spectrum obtained with the ARCS time-of-flight spectrometer at the SNS at $T = 16$ K. Blank patches are due to gaps between detectors. Since the chopper spectrometer allows only two components of $Q$ to be independent and the third is a function of energy, in this pattern the value of $L$ varies with energy from $L =$ -2.5 at 5 meV to $L = 3.5$ at 30 meV. The pattern shown here extends up to about 25 meV, forming a column of excitations that is narrow in $H$ and $K$ but quite broad in energy and $L$. To characterize the intensity of the column at [0.5, 0.5, 2] we took the difference between the intensities of the HFIR data at $H = 0.5$ and 0.6, and defined this value as the column intensity. As shown in Fig. 5, at $T = 30$ K the column intensity is nearly constant up to 20 meV, consistent with the ARCS data in Fig. 4. The difference between the data at $T = 16$ K and 30 K in Fig. 5 corresponds to the results in Fig. 1, except that here the data quality was compromised by the noise from the data at $H = 0.6$.

In addition, the intensity of the column was measured at two energies ($E = 10$ meV and 14 meV) and at several temperatures ($T = 16$K, 30K, 50K, 200K) for $Q = [0.5, 0.5, 2]$. The column intensity was defined again as the difference between the intensities at [0.5, 0.5, 2] and at [0.6, 0.6, 2]. The data, corrected for the Bose-Einstein (B-E) factor, are plotted in Fig. 6. The data clearly show that the excitation persists to high temperature. At $T = 200$K the intensity is still nearly 25% of the value at $T = 30$ K for both of the measured energies after the B-E correction. This implies that the energy scale for the thermal reduction in intensity is on the order of 100 K.

It is widely believed that the electronic state of the FeAs based compounds can be described well by an itinerant fermion picture[12]. Indeed, the density functional theory (DFT) calculations generally agree well with the observed electronic structure and properties of the parent compounds[13]. But the DFT fails with respect to magnetism. In the first place, DFT overestimates the magnetic moment of the parent compounds. In addition, for the compositions which exhibit superconductivity it universally predicts a magnetic ground state, whereas they are generally paramagnetic above the superconducting critical temperature, $T_C$[14, 15].

The data shown here offer some suggestions for this conundrum. First, the observed spin excitation spectrum is too sharp in the [$H, K$] plane to be compatible with the bare susceptibility calculated from the band structure. The diameter of the Fermi surfaces in the [$H, K$] plane, both for holes and electrons, are of the order of 0.3 Å$^{-1}$, whereas the spin excitation at 10 meV is only 0.16 Å$^{-1}$ in width. Second, the intensity of the spin excitation decreases with increasing temperature, while the bare electronic susceptibility is independent of temperature, except for the Bose factor. These results imply that the observed spin susceptibility, $\chi''(Q, \omega)$, is strongly enhanced by the exchange interaction, for instance by $\chi''(Q, \omega) = \chi_0''(Q, \omega)/(1 - \sigma(Q, \omega) \chi_0''(Q, \omega))$, where $\chi_0''$ is the bare susceptibility and $\sigma$ is the exchange enhancement factor. The exchange enhancement would lead to spin correlations, which narrow the excitation spectrum, and the enhancement would decrease with increasing temperature[16]. Both of these expected consequences are consistent with the observations here. The fact that the narrowing of the spectrum is confined to the [H, K] plane suggests that the spin correlations are strongly two-dimensional. This is rather surprising, since the spin correlations in the undoped samples are much more three-dimensional[17, 18, 19, 20]. The energy scale we determined for the temperature dependence, ~10 meV, is significantly smaller than the spin wave stiffness, which is consistent with $J \sim 70$ meV. This implies that electrons in the system are far from the localized spin limit that can be described by the fixed real space exchange constant, but are closer to the itinerant limit[16].

It is possible that the electron spins are locally polarized as suggested by core-level spectroscopy[21] and LDA calculations[22, 23], while the two-dimensional spin fluctuations are suppressing the long-range static order. Thus even though the FeAs compounds are basically itinerant electron systems, the exchange enhancement gives rise to a partially localized nature. On the other hand the undoped cuprates are Mott insulators in which spins are totally localized, but doping gives some itinerant character to charge carriers, particularly oxygen holes. For this reason whereas the undoped Fe pnictides and the cuprates are fundamentally different in the nature of charge carriers, doping may make them more similar than they appear. It is unclear, however, whether this similarity extends to the mechanism of superconductivity or not.

The authors are grateful to D. J. Singh, I. I. Mazin, T. Yildirim, N. Mannella and D. J. Scalapino for stimulating and useful discussions. The work at the University of Tennessee was supported by the Department of Energy EPSCoR Implementation award, DE-FG02-08ER46528. The work at the Oak Ridge National Laboratory was supported by the Scientific User Facilities Division and by the Division of Materials Science and Engineering, Office of Basic Energy Sciences, Department of Energy.

**Figure captions:**

Fig. 1   The difference between the inelastic scattering intensities taken at two temperatures, $I(T = 16$ K$) - I(T = 30$ K$)$, at the momentum exchange, $\bm{Q} = (0.5, 0.5, 0)$ and $(0.5, 0.5, 2)$.

Fig. 2   The difference between the inelastic scattering intensities taken at two temperatures, $I(T = 16$ K$) - I(T = 30$ K$)$, at $\bm{Q} = (2.5, 2.5, 0)$, suggesting that the phonon contribution to the data in Fig. 1 is negligibly small.

Fig. 3   The difference in the intensities for the $L$-scans at $\bm{Q} = (0.5, 0.5, L)$, $E = 12$ meV measured at $T = 16$ K and at 30 K.  The solid line is the square of the $Fe^{2+}$ spin-only magnetic form factor.

Fig. 4  The scattering intensity map in the $H + K = 1$, $E$ plane, obtained with the ARCS spectrometer of the SNS.  $L$ index is a function of energy, and varies from $L = -1.5$ at $E = 5$ meV to $L = 3$ at $E = 25$ meV.  The L-index at the center, corresponding to [0.5, 0.5, L], is shown on the right.  Blank spots are due to gaps between detectors.  The intensities around [0, 1, L], $E = 11$-17 meV and around the edges are artifacts due to the ends of the detectors.

Fig. 5  Energy dependence of the height of the column intensity defined by the difference between the intensities at $Q = [0.5, 0.5, 2]$ and $[0.6, 0.6, 2]$, measured at $T = 16$ and 30K.  The difference between these scans constitutes the resonance peak shown in Fig. 1.

Fig. 6  Temperature dependence of the height of the column intensity, defined by the difference between the intensities at $Q = [0.5, 0.5, 2]$ and $[0.6, 0.6, 2]$, measured at $E = 10$ and 14 meV, corrected for the Bose-Einstein factor.

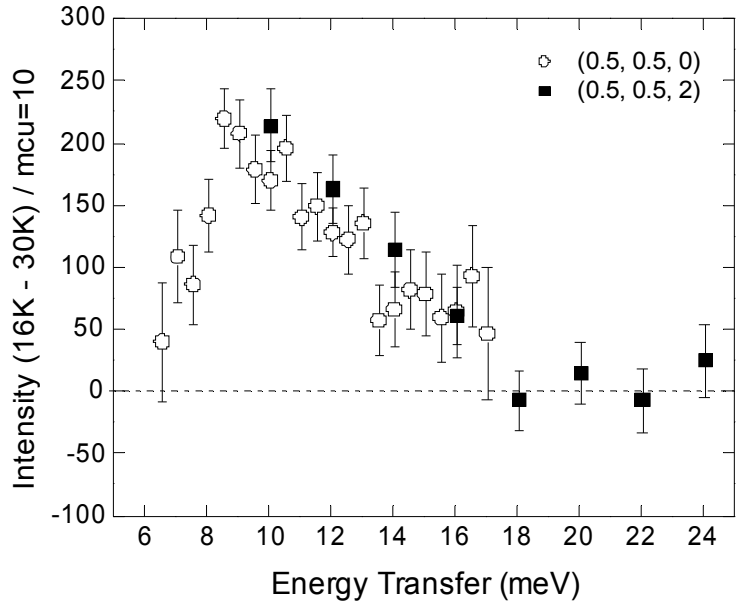

Fig. 1

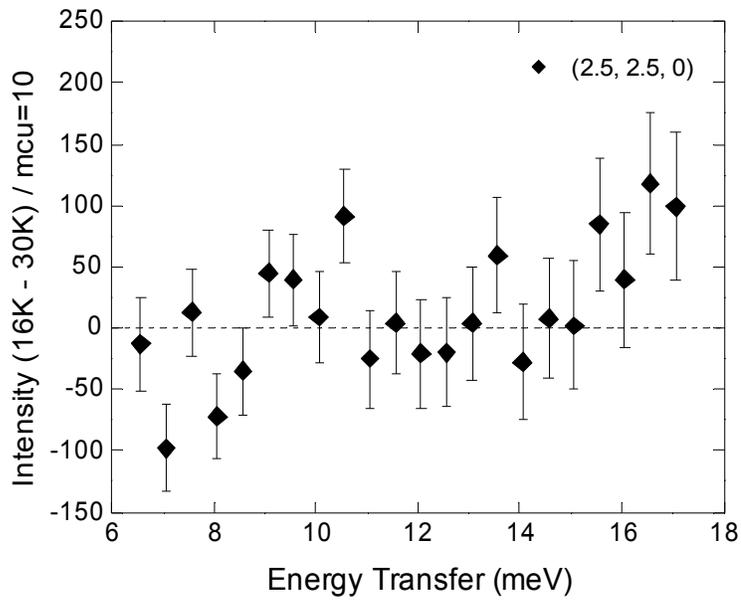

Fig. 2

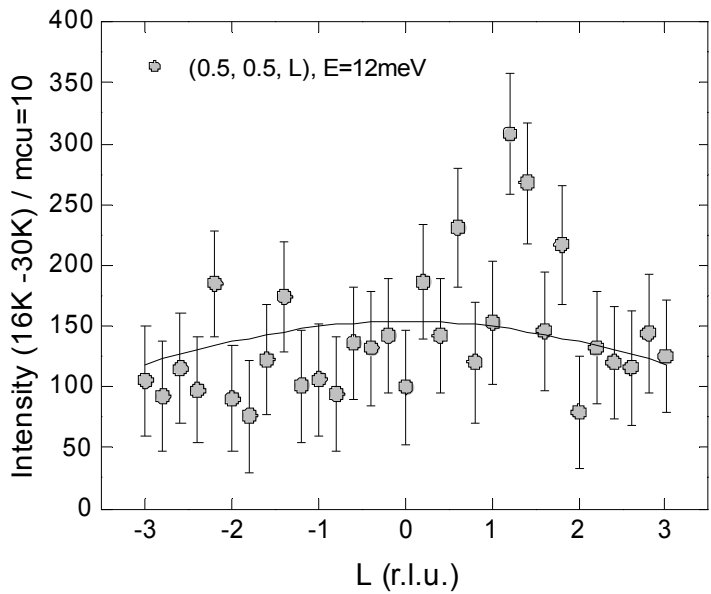

Fig. 3

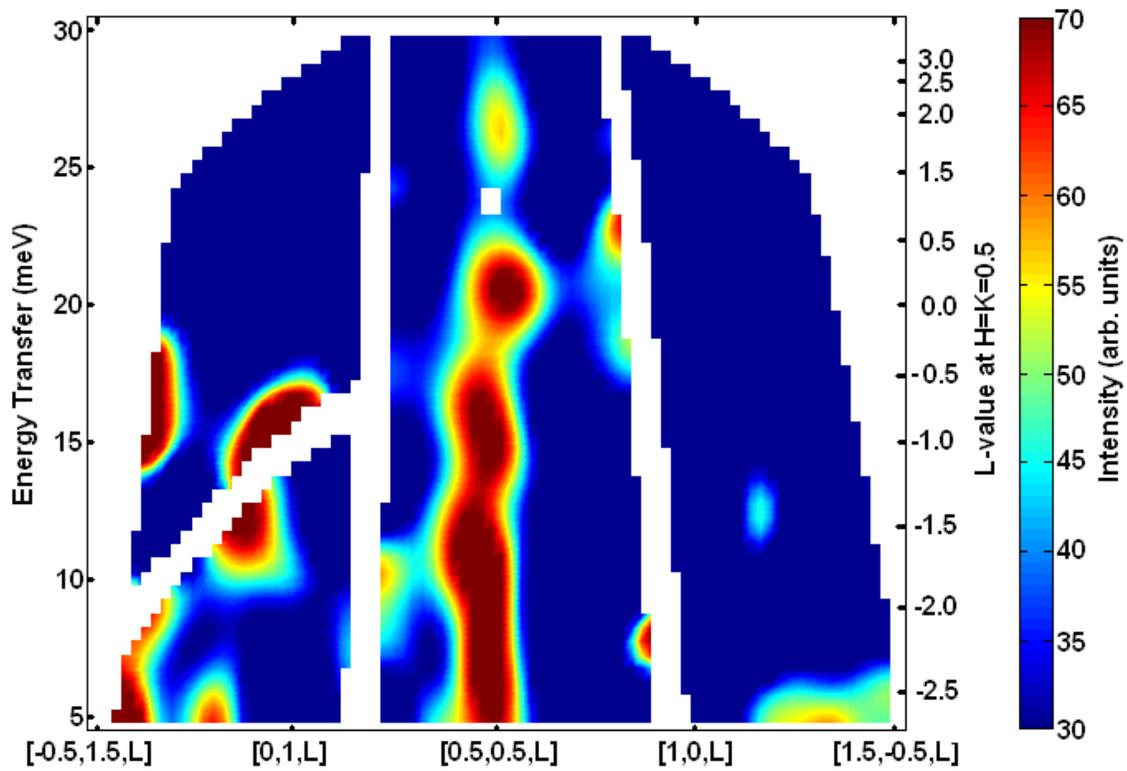

Fig. 4 (color)

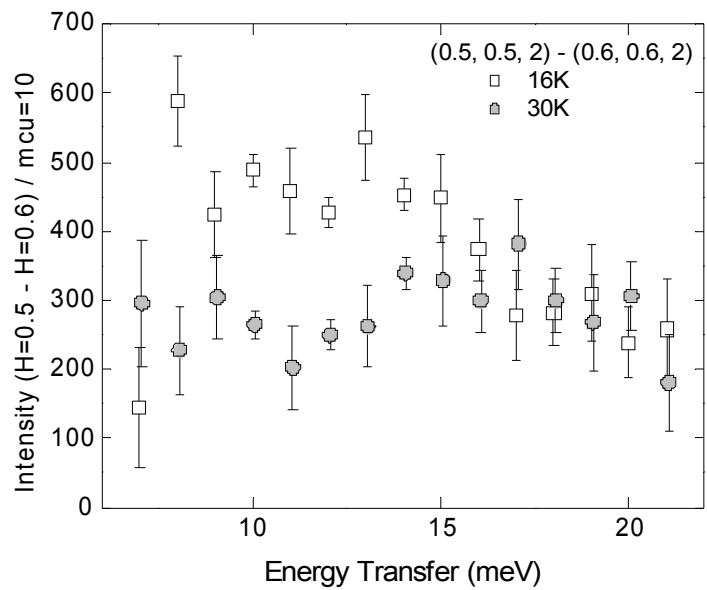

Fig. 5

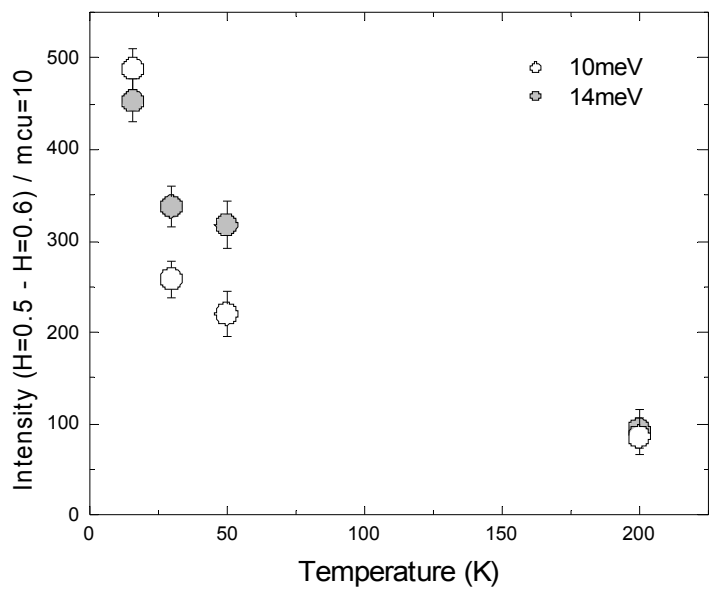

Fig. 6